\documentclass[
  aps,%
  12pt,%
  final,%
  notitlepage,%
  oneside,%
  twocolumn,%
  nobibnotes,%
  nofootinbib,%
  superscriptaddress,%
  noshowpacs,%
  centertags]%
          {revtex4}
\usepackage{amssymb}
\usepackage{amsmath}
\usepackage{longtable}
\usepackage{graphicx}
\usepackage{psfrag}
\usepackage{aas_macros}

\begin{document}
\title{Evolution of Galaxies and the Tully--Fisher Relation}
\author{\firstname{E.~P.}~\surname{Kurbatov}}
\email{kurbatov@inasan.ru}
\affiliation{Institute of Astronomy, Moscow, Russia}
\author{\firstname{A.~V.}~\surname{Tutukov}}
\email{atutukov@inasan.ru}
\affiliation{Institute of Astronomy, Moscow, Russia}
\author{\firstname{B.~M.}~\surname{Shustov}}
\email{bshustov@inasan.ru}
\affiliation{Institute of Astronomy, Moscow, Russia}
\begin{abstract}

We study the evolution of the [O/Fe]--[Fe/H] relation and the dependence
of the iron abundance on distance from the galactic plane $z$ in a one-zone
model for a disk galaxy, starting from the beginning of
star formation \cite{Wiebe98:AZ-75-1-3}. We obtain good agreement with the
observational data, including, for the first time, agreement for the
[Fe/H]\mbox{--}$z$ relation out to heights of 16~kpc. We also study the
influence of the presence of dark matter in the galaxies on the star-formation
rate. Comparison of the observed luminosity of the Galaxy with the model
prediction places constraints on the fractional mass of dark matter, which
cannot be much larger than the fractional mass of visible matter, at least
within the assumed radius of the Galaxy, $\sim 20$~kpc. We studied the
evolution of disk galaxies with various masses, which should obey the
Tully--Fisher relation, $M\propto R^2$. The Tully--Fisher relation can be
explained as a combination of a selection effect related to the observed
surface brightnesses of galaxies with large radii and the conditions for the
formation for elliptical galaxies.

\end{abstract}

\maketitle
\section{Introduction}

Many thousands of theoretical papers have been written on the evolution of
galaxies, often applying sophisticated mathematical methods: $N$-body
models, multidimensional gas-dynamical models, statistical methods etc.
However, a number of very important results can be obtained using methods that
are relatively simple and are not computationally demanding.
In \cite{Firmani92:AAP-264-37,Wiebe98:AZ-75-1-3,Firmani94:AAP-288-713,Wiebe99:AAP-345-93,Wiebe:AZ},
we suggested and implemented an approach that is both simple mathematically and
reasonably self-consistent in the treatment of all complex physical processes
determining the evolution of a galaxy.

In mathematical terms, this evolution reduces to the solution of two equations.
One defines the star-formation rate (SFR) assuming the complete ionization
of the gas component of the galaxy, which is taken to be homogeneously
distributed over the galactic disk. The other equation describes variations
of the thickness of the galactic disk based on the condition of virial
equilibrium, the input of mechanical energy into the gas component by
supernovae, and the dissipation of energy due to collisions of gas clouds.
Numerical modeling must also take into account the evolution of the galactic
stellar component: the return of matter to the interstellar medium by old stars
after the formation of the endproducts of their evolution --- black holes,
neutron stars, and degenerate dwarfs --- and the exchange of matter between
the galaxy and intergalactic medium.

Despite the simplicity of this approach, it led to a fairly complete picture
of the evolution of disk galaxies --- in particular, the Milky Way --- that
was consistent with observations. This made it possible for the first time
to distinguish the role of the loss of heavy elements from the Galaxy during
the formation of the radial chemical-composition gradient, and to derive
self-consistent distributions of metals in the $z$ direction over a scale
of $\sim 2$~kpc. This provided an explanation for the origin of enhanced
metallicities (by a factor of a few over the solar value) of galaxies harboring
quasars in their nuclei, via the enhanced density and higher minimum mass
of the stars in these regions \cite{Wiebe:AZ}.

\begin{figure*}
  \psfrag{[O/Fe]}{$\rm [O/Fe]$}
  \psfrag{[Fe/H]}{$\rm [Fe/H]$}
  \psfrag{10^7}[l][l]{$\tau_{\rm a} = 10^7$ yr}
  \psfrag{10^8}[l][l]{$\tau_{\rm a} = 10^8$ yr}
  \psfrag{10^9}[l][l]{$\tau_{\rm a} = 10^9$ yr}
  \includegraphics[width=15cm]{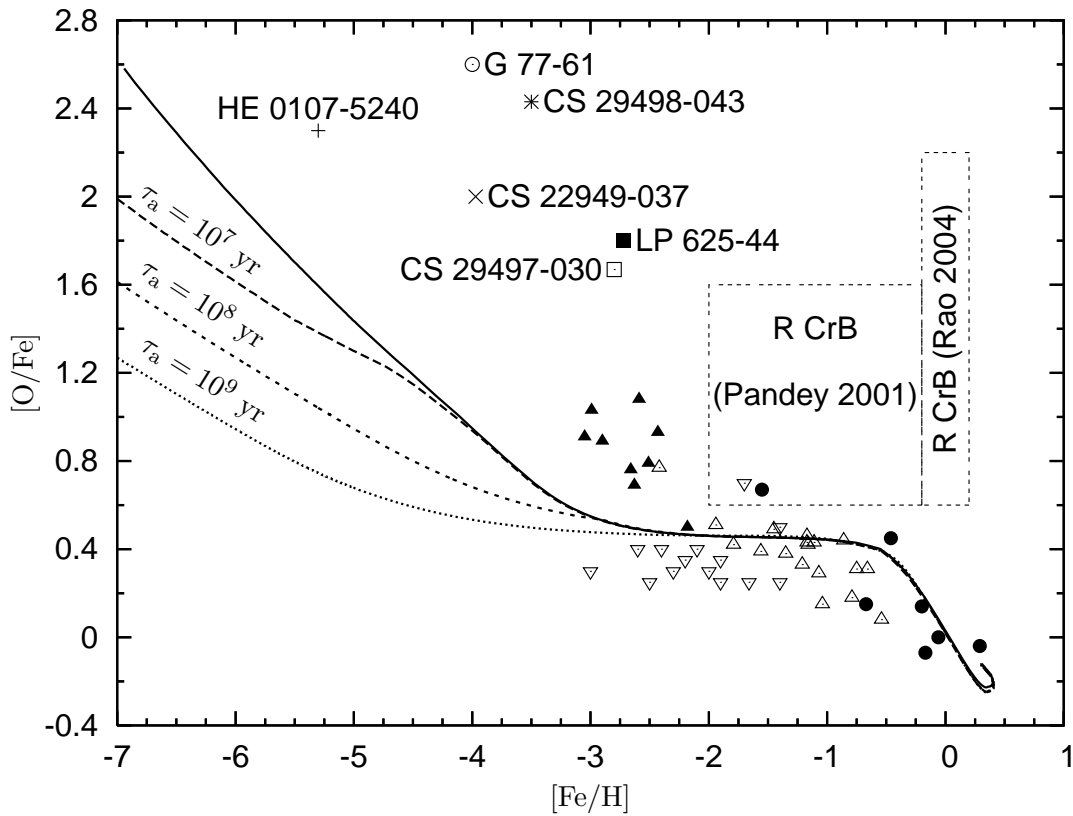}
  \caption{[O/Fe]--[Fe/H] dependence for models with various initial
    accretion times $\tau_{\textrm{a}}$. The solid curve shows the standard
    model ($\tau_{\textrm{a}} = 0$). The filled circles show data on
    subgiants \cite{Thoren04:AAP-425-187}, open triangles data on subgiants and
    main-sequence stars \cite{Nissen02:AAP-390-235}, filled triangles data on
    subdwarfs \cite{Israelian01:APJ-551-833}, inverted open triangles data on
    halo giants \cite{Barbuy88:AAP-191-121}, and rectangles the positions of
    R~CrB stars \cite{Pandey01:MNRAS-324-937,Rao:astro-ph:0410648}. The
    remaining objects are HE~0107-5240 \cite{Bessel04:APJL-612-L61},
    CS~22949-037 \cite{Cayrel04:AAP-416-1117},
    CS~29498-043 \cite{Israelian04:AAP-419-1095},
    CS~29497-030 \cite{Israelian04:AAP-419-1095},
    LP~625-44 \cite{Israelian04:AAP-419-1095}, and
    G~77-61 \cite{Plez:astro-ph:0501535}.
    \hfill}
  \label{fig:ofe-feh}
\end{figure*}

The present paper continues our study of the evolution of disk galaxies
using the model developed by us earlier. This model enabled us to trace
the chemical evolution of galaxies starting from their formation. However,
the absence of data on very old, low-metallicity stars prevented comparison
of the model results with observational data for the initial stages of
the evolution of the Galaxy.  New data for stars with extremely low
metallicities has recently become available, making it possible to trace
the early stages of the enrichment of the Galaxy in heavy elements. Section~2
discusses the evolution of abundances implied by our theoretical modeling.
Section~3 discusses modern data on the distribution of metals in the direction
of the Galactic rotational axis on scales of 16~kpc. In our previous studies,
we used only fairly old data on stars that are located not further than 3~kpc
from the disk --- again hindering application of the model to the earliest
stages of Galactic evolution. Another important circumstance influencing
the evolution of the Galaxy is the presence of dark matter in the Galaxy.
The nature and properties of dark matter are far from fully understood (see,
e.g., the review \cite{Shustov05:ISSC-207}), but the necessity of taking this
factor into account is beyond doubt. Section~4 presents an estimate of the
fractional mass of dark matter obtained in our model.

The good agreement of the model results with the observations encourages us
to use the model to interpret various statistical relations (mass--metallicity,
mass--luminosity, etc.). In recent years, possibilities for applying the
Tully--Fisher relation \cite{Tully77:AAP-54-661}, which was discovered in 1977
by the two named authors, in various studies of galaxies have been widely
discussed. This law relates the luminosity and rotational velocity of a disk
galaxy. There are various ideas about the origin of the Tully--Fisher
relation. In Section~5, we suggest an explanation that seems to us quite
natural, based on an analysis of galactic evolution using our model.  Finally,
Section~6 presents a summary of our results.

We made some modifications to the set of standard parameters for the Galaxy
for the computations; these are described in the corresponding Sections.

\section{Evolution of the oxygen and iron abundances}

Analyses of the oxygen and iron abundances are the classical methods for
testing our understanding of the evolution of stars and galaxies. These
elements were adopted as tools for directly analyzing the abundances of the
products of type~I supernovae (SN~Ia, which produce mainly iron) and type~II
supernovae (SN~II or SN~Ib,c, which produce mainly oxygen). The time
between the formation of a massive star and its explosion as a SN~II or
SN~Ib,c does not exceed $\sim 10^7$~yr. According to one possible scenario,
SN~Ia result from the mergers of degenerate
dwarfs \cite{Iben84:APJS-54-335}. The lifetimes of SN~Ia precursors cover a
wide range, from $10^7$ to $10^{10}$~yr, with the most probable values being
from $2\times10^8$~yr to $2\times10^9$~yr \cite{Tutukov02:AZ-79-1}. The
possibility that SN~Ia make a significant contribution to the production of
iron in the Galaxy at ages lower than $\sim 10^9$~yr is supported by the
discovery of an [O/Fe] ratio equal to the solar value in a distant
quasar \cite{Barth03:APJL-594-L95} with $z = 6.4$ (age $\sim
7\times10^8$~yr). We have used here a standard model for the evolution of the
Galaxy: the closed model with outer radius $R = 20$~kpc and mass  $M =
2\times10^{11} M_\odot$ described in \cite{Wiebe98:AZ-75-1-3}. The only
difference is in the lifetime of SN~Ia precursors (which was assumed to be
$10^9$~yr, close to the median value in the scenario for the evolution of close
binaries \cite{Tutukov02:AZ-79-1}). In previous computations of the standard
model \cite{Wiebe98:AZ-75-1-3}, this time was taken to be
$3\times10^8$~yr. Test runs have shown that this change does not result in any
fundamental changes in the results of the computations; therefore, we shall
adopt this modified model as our standard.

We considered the evolution of the [O/Fe]\mbox{--}[Fe/H] relation in a number
of previous studies \cite{Wiebe98:AZ-75-1-3,Wiebe99:AAP-345-93,Wiebe:AZ}, using
stars in the solar neighborhood with relative iron abundances [Fe/H] $ >
-3$. The metallicity range considered here is wider, extending to
$\textrm{[Fe/H]} = -7$. The main results are presented in
Fig.~\ref{fig:ofe-feh}. When $\textrm{[Fe/H]} < -1$, the abundances of oxygen
and iron are determined primarily by SN~II explosions, leading to an
overabundance of oxygen by a factor of two to four. At an age of $\sim
5\times10^8$~yr, SN~Ia begin to explode, producing primarily
iron \cite{Bensby04:AAP-415-155} (in the closed model, this age corresponds to
[Fe/H] $\approx -0.5$). The ratio [O/Fe] begins to decline with time (along
with the increase of $\textrm{[Fe/H]}$), and approaches the solar value at an
age of several Gyr. The model provides an acceptable distribution for most
stars with $\textrm{[Fe/H]}> -3$.

Among stars with the solar abundance of heavy elements, it has long been known
that R~CrB stars have abnormally high abundances of oxygen and carbon. The
positions of R~CrB stars with solar iron abundances are plotted in
Fig.~\ref{fig:ofe-feh}, based on the data of \cite{Pandey01:MNRAS-324-937};
low-metallicity ($\textrm{[Fe/H]}\approx -0.5\text{--}-2.0$) R~CrB stars are
plotted based on the data of \cite{Rao:astro-ph:0410648}. The overabundance of
oxygen in R~CrB stars reaches nearly two orders of magnitude, and may be nearly
independent of [Fe/H] (Fig.~\ref{fig:ofe-feh}). The reason for the
overabundance of O and C in R~CrB stars is the episodic penetration of the
convective envelope of the star on the asymptotic giant branch onto the layer
enriched in the products of helium burning \cite{Iben83:ARAA-54-271}. This
process enhances the observed abundances of C, O, and N in red supergiants by
nearly two orders of magnitude.  This oxygen-enrichment mechanism is possible
only for red (super)giants  with luminosities that are more than two orders of
magnitude higher than the solar luminosity. Stars of lower mass may become
enriched in carbon or oxygen due to mass exchange between the components of
close binaries. This may explain the origin of such oxygen-rich stars as
LP~625-44 and CS~29497-030 \cite{Israelian04:AAP-419-1095}. Stars with carbon
abundances $\textrm{[C/Fe]}= 2$ and [Fe/H] $= -4$ are known
\cite{Ryan:astro-ph:0211608}.

An increasing number of observations of stars with $\textrm{[Fe/H]}< -3$ have
been reported in recent years. The very low metallicities of these stars
reflect the earliest stages of the evolution of the Galaxy
\cite{Suda04:APJ-611-476}.  Such stars can be arbitrarily separated into two
groups (Fig.~\ref{fig:ofe-feh}). The first group contains stars that show a
substantial excess of oxygen relative to the model: G~77-61
\cite{Plez:astro-ph:0501535}, CS~29498-043 \cite{Israelian04:AAP-419-1095},
CS~22949-037 \cite{Cayrel04:AAP-416-1117}, CS~22957-027
\cite{Umeda05:APJ-619-445} (${\textrm{[O/Fe]}} \approx 2$, ${\textrm{[Fe/H]}} =
-3.11$), and CS~31062-012 \cite{Umeda05:APJ-619-445} (${\textrm{[O/Fe]}}
\approx 2$, ${\textrm{[Fe/H]}} = -2.55$), and R~CrB.  The second group contains
the giant HE~0107-5240 \cite{Bessel04:APJL-612-L61}. While the $\textrm{O}$
excesses in stars of the first group can be explained by the penetration of
their convective envelopes into the layer enriched in oxygen or by mass
exchange between binary components, of the oxygen overabundance in the stars
with the lowest metallicities may be primordial, indicating that they formed
during the first $\sim 10^7$~yr of the Galaxy's existence. Figure~1 shows that
the [O/Fe] ratio in the earliest stages of the Galaxy's evolution is very high,
and decreases with age. Since SN~Ia have not yet started to enrich the
interstellar medium in iron, a natural question arises: What is the reason for
this behaviour of [O/Fe]?

The reason for the increase of the [O/Fe] ratio is the increase in the
production of oxygen and the decrease in the production of iron with increasing
initial masses of SN~II precursors. According to the numerical models
\cite{Maeder92:AAP-264-105,Tielemann96:APJ-460-408}, the amount of oxygen
$M_{\textrm{O}}$ produced by stars with masses of 15--120~$M_\odot$ can be
written
\begin{gather*}
  \frac{M_{\textrm{O}}}{M_\odot} = 0.01 \left(
  \frac{M_{\textrm{i}}}{M_\odot} \right)^{1.7},
\end{gather*}
where $M_{\textrm{i}}$ is the initial mass of the star. The amount of iron
produced by 13--25~$M_{\odot}$ stars can be estimated as
\cite{Tielemann96:APJ-460-408}
\begin{gather*}
  \frac{M_{\textrm{Fe}}}{M_\odot} =
  21 \left( \frac{M_{\textrm{i}}}{M_\odot} \right)^{-1.87}.
\end{gather*}
Combining these relations, we obtain
\begin{gather}
  \frac{M_{\textrm{O}}}{M_{\textrm{Fe}}}
  \approx 5\times10^{-4} \left( \frac{M_{\textrm{i}}}{M_\odot}
  \right)^{3.57}.
  \label{eq:Modelled_ofe}
\end{gather}
This expression is valid for $13 < M_{\textrm{i}}/M_{\odot} <120$, since only
stars of these masses were studied in
\cite{Maeder92:AAP-264-105,Tielemann96:APJ-460-408}. Of course, when working
with~(\ref{eq:Modelled_ofe}), we must bear in mind that, while the production
of oxygen by massive stars is estimated relatively accurately, the iron yield
remains uncertain. It may depend on several parameters that are not accurately
known, such as the initial chemical composition of the stars, the rotational
velocity of the presupernova core, the magnetic-field strength etc. It is clear
from observations that the iron yield may be even smaller than the models
\cite{Tielemann96:APJ-460-408} predict. For instance, SN~1997D, with a total
mass of $\sim 25M_{\odot}$, ejected only $2\times10^{-3} M_{\odot}$ of nickel
(which produces iron as it decays) \cite{Umeda03:NAT-422-871}, much less than
the $\sim 0.05M_{\odot}$ expected from the models
\cite{Tielemann96:APJ-460-408}. On the other hand, SN~2002ap, with an initial
mass of $\sim 25 M_\odot$, produced $0.07 M_{\odot}$ of nickel
\cite{Mazzali02:APJL-572-L61}, close to the model expectations.

According to~(\ref{eq:Modelled_ofe}), obtaining the ratio ${\textrm{[O/Fe]}}
\approx 2$ observed in the extreme low-metallicity star HE~0107-5240
(Fig.~\ref{fig:ofe-feh}) requires a supernova with an initial mass of  $\sim 50
M_{\odot}$ and an age of $3\times10^6$~yr. Thus, judging from the iron
abundance, HE~0107-5240 is among the first stars to be formed in the Galaxy,
during the first several million years of its existence. Stars with
${\textrm{[Fe/H]}} \lesssim -3$ have masses of the order of a solar mass. This
suggests that the initial enrichment of the Galaxy in heavy elements was
accomplished by ordinary stars with masses from $\sim 1 M_{\odot}$ to $\sim 100
M_{\odot}$.

The metallicity of the first Galactic stars could grow due to the accretion of
interstellar gas. If the radius for capturing interstellar gas by a star is
$r = 2 G M v^{-2}$, where $G$ is the gravitational constant,  $M$ the mass of
the star, and $v$ its relative velocity, the star can accrete during the
lifetime of the Galaxy a mass
\begin{gather}
  \frac{\Delta M}{M_\odot} \approx 10^{-5}\frac{n_{\textrm{H}}}{v_{30}^3},
  \label{eq:Accretion_Rate}
\end{gather}
where  $n_{\textrm{H}}$ is the current number density of the hydrogen in the
gaseous phase (in $cm^{-3}$) and $v$ is the relative velocity (in units of
30~km/s). According to~(\ref{eq:Accretion_Rate}), even in stars that are
completely devoid of heavy elements, the relative iron abundance could become
equal to that observed in HE~0107-5240 (Fig.~\ref{fig:ofe-feh}) due to
accretion (if there is a negligible role of the stellar wind, low mixing
efficiency in the outer layers of the star, etc.). Note that low-metallicity
stars could have initially belonged to a former low-mass satellite that merged
with the Galaxy in the past. The evolution of the chemical composition of
low-mass spheroidal galaxies may end in an early phase, after loss of the
gaseous component. Naturally, stars in such a galaxy could retain their high
[O/Fe] values.

\begin{figure*}
  \psfrag{[Fe/H]}{$\rm [Fe/H]$}
  \psfrag{z}{$z$, kpc}
  \includegraphics[width=15cm]{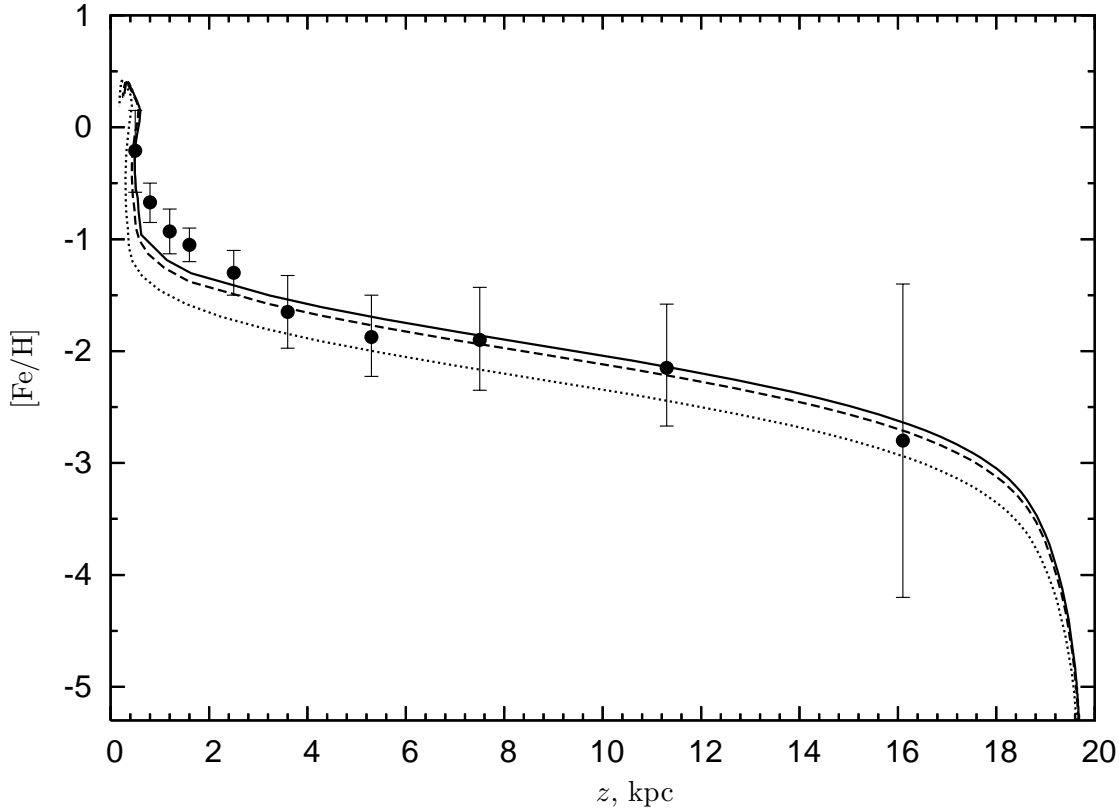}
  \caption{Distribution of iron abundance with height above the Galactic plane
    in the standard model (solid curve) and in a model taking into account the
    loss of heavy elements due to SN~II explosions, for fractional masses of
    heavy elements lost equal to 0.16 (dashed curve) and 0.5 (dotted
    curve). The solid circles show the data from \cite{Du04:AJ-128-2265}.
    \hfill}
  \label{fig:feh-z}
\end{figure*}

For completeness, we note an additional possible means of enrichment of
population-III low-mass stars (which are initially devoid of heavy elements)
in metals.  Most stars are formed in stellar clusters, which, as a rule, are
disrupted almost immediately after their formation because of the loss of the
gaseous component due to the formation of HII regions and supernova
explosions \cite{Tutukov78:AAP-70-57}. The first stars may also have formed in
clusters. Then, some fraction of the lost gas, along with the products of the
first, most massive, supernovae, would be captured by low-mass stars with the
initial low-metal chemical composition. The capture radius is $r\approx
2GM_{*}v_{\textrm{ex}}^{-2}$, where $M_{*} \approx M_{\odot}$ is the mass of
the cluster star and $v_{\textrm{ex}}$ the velocity of the expanding gaseous
envelope of the cluster. The fraction of the captured matter will then be
\begin{gather}
  \alpha = \frac{G^2 M_{*}^2}{v_{\textrm{ex}}^4 R^2} = \left(
  \frac{v_{\textrm{ff}}}{v_{\textrm{ex}}} \right)^4 N^{-2},
  \label{eq:Accretion_Rate_In_Cluster}
\end{gather}
where $R$ is the cluster radius, $v_{\textrm{ff}}$ the free-fall velocity at
the edge of the cluster, and $N$ the number of stars in the cluster (assuming
that all have the solar mass). If we assume that
$v_{\textrm{ff}} = v_{\textrm{ex}}$, $N = 10^3$ and $M_{\textrm{Fe}} \approx
0.01~M_\odot$ ($M_{\textrm{O}} \approx 50~M_{\odot}$
\cite{Tielemann96:APJ-460-408})), then a solar-mass star can capture $\sim
10^{-8} M_{\odot}$ of iron. This produces a star with an iron abundance of
$\textrm{[Fe/H]} \approx -5$, similar to that observed in HE~0107-5240
(Fig.~\ref{fig:ofe-feh}).  It is obvious that a high oxygen abundance will be
observed, since stars with masses exceeding $\sim 50 M_{\odot}$ have
$\textrm{[O/Fe]} \approx 2$, as was shown above. Note that not all the
parameters in~(\ref{eq:Accretion_Rate_In_Cluster}) are well enough known to
enable confident conclusions about the role of this mechanism in the enrichment
of population-III stars in heavy elements. The accretion of interstellar gas
enriched in heavy elements produced by the first stars may substantially
complicate, or even make impossible, the identification of low-mass
population-III stars.

Note that the [O/Fe]--[Fe/H] evolutionary curve obtained for the closed model
shown in Fig.~\ref{fig:ofe-feh} represents an upper bound for the oxygen
abundance. In low-mass disk galaxies, the products of SN~II explosions can
efficiently leave the parent galaxies via the formation of supershells
\cite{Igumenshchev90:AAP-30-524}, thereby reducing the growth rate of the
oxygen abundance in these galaxies. Since type~Ia supernovae do not produce
supershells, the iron abundance in low-mass disk galaxies increases at the
usual pace. The example provided by the galaxy IZw~18
\cite{Aloisi03:APJ-595-760}, which has active star formation with
$\textrm{[Fe/H]} = -1.76 \pm 0.12$ and $\textrm{[O/Fe]} = -0.3 \pm 0.3$, shows
that the loss of the products of type~II supernovae may, indeed, be very
important for dwarf galaxies. As our model computations show, allowing for the
loss of type~I supernova products in massive galaxies does not lead to any
substantial differences in the [O/Fe]--[Fe/H] distribution. Moreover, special
runs of the model taking into account the finite time for the accumulation of
the Galactic matter show that increasing the Galaxy's formation time
substantially reduces the [O/Fe] ratio at low [Fe/H]
(Fig.~\ref{fig:ofe-feh}). Increasing the amount of data on low-metallicity
stars in the future may enable estimation of the time for the accumulation of
the Galaxy's mass.

\section{Distribution of heavy elements above the galactic disk}

Let us now consider the distribution of heavy elements with height $z$ above
the Galactic disk. Various observations show that the metallicity gradient
with $z$ measured using field stars may reach $-0.65$/kpc (according to data
for the thin disk \cite{Yamagata94:IAUS-161-420})), while the lowest value
obtained from observations of open clusters is $-0.34$/kpc
\cite{Carraro98:MNRAS-296-1045}. At the same time, some observations of field
and open-cluster stars show no evidence for a metallicity gradient
\cite{Gilmore85:AJ-90-2015,Friel95:ARAA-33-381}. To summarize, at $z < 4$~kpc,
the gradient is in the range $(-0.55\ldots-0.33)$/kpc, while it is in the range
$(-0.18\ldots 0)$/kpc at larger $z$ \cite{Du04:AJ-128-2265}.

\begin{table*}
  \caption{Evolution of the main parameters of the standard Galaxy model with
    age (SFR is the star formation rate in $M_{\odot}/$yr, $H_\mathrm{g}$ the
    semi-thickness of the gaseous disk, $M_\mathrm{g}$ the mass of the gas,
    $Z$ the fractional mass of heavy elements, $\tau_\mathrm{dust}$ the optical
    depth of the dust, $L_\mathrm{opt} = L_\mathrm{bol}/\tau_\mathrm{dust}(1 -
    e^{-\tau_\mathrm{dust}})$ the optical luminosity with allowance for
    absorption by dust)}
  \bigskip
  \begin{tabular}{c|c|c|c|c|c|c|c|c|c}
    \hline
    $t$, yr & \multicolumn{1}{c|}{SFR} & $H_\mathrm{g}$, pc &
    $M_\mathrm{g}/M_{\odot}$ & $Z$ & \multicolumn{1}{c|}{[O/H]} &
    \multicolumn{1}{c|}{[Fe/H]} &
    \multicolumn{1}{c|}{$\tau_\mathrm{dust}$} & $L_\mathrm{opt}/L_{\odot}$
    &$L_\mathrm{bol}/L_\odot$ \\
    \hline
$3.6\cdot10^{6}$ & $34.3$ & $2.0\cdot10^{4}$ & $2.0\cdot10^{11}$ &
 $7.0\cdot10^{-7}$ & $-4.4$ & $-6.9$ & $0.0012$ & $1.1\cdot10^{10}$ &
 $1.1\cdot10^{10}$ \\
$4.8\cdot10^{6}$ & $34.4$ & $2.0\cdot10^{4}$ & $2.0\cdot10^{11}$ &
 $2.0\cdot10^{-6}$ & $-3.9$ & $-5.8$ & $0.0034$ & $1.2\cdot10^{10}$ &
 $1.2\cdot10^{10}$ \\
$6.3\cdot10^{6}$ & $34.5$ & $2.0\cdot10^{4}$ & $2.0\cdot10^{11}$ &
 $4.3\cdot10^{-6}$ & $-3.6$ & $-5.1$ & $0.0072$ & $1.3\cdot10^{10}$ &
 $1.3\cdot10^{10}$ \\
$8.2\cdot10^{6}$ & $34.7$ & $2.0\cdot10^{4}$ & $2.0\cdot10^{11}$ &
 $7.5\cdot10^{-6}$ & $-3.4$ & $-4.7$ & $0.0012$ & $1.3\cdot10^{10}$ &
 $1.4\cdot10^{10}$ \\
$1.0\cdot10^{7}$ & $34.9$ & $1.9\cdot10^{4}$ & $2.0\cdot10^{11}$ &
 $1.2\cdot10^{-5}$ & $-3.2$ & $-4.3$ & $0.020$ & $1.4\cdot10^{10}$ &
 $1.4\cdot10^{10}$ \\
$1.5\cdot10^{7}$ & $35.3$ & $1.9\cdot10^{4}$ & $2.0\cdot10^{11}$ &
 $2.2\cdot10^{-5}$ & $-3.0$ & $-3.8$ & $0.036$ & $1.5\cdot10^{10}$ &
 $1.6\cdot10^{10}$ \\
$2.5\cdot10^{7}$ & $36.2$ & $1.9\cdot10^{4}$ & $2.0\cdot10^{11}$ &
 $4.3\cdot10^{-5}$ & $-2.7$ & $-3.3$ & $0.071$ & $1.7\cdot10^{10}$ &
 $1.7\cdot10^{10}$ \\
$3.4\cdot10^{7}$ & $37.0$ & $1.8\cdot10^{4}$ & $2.0\cdot10^{11}$ &
 $6.3\cdot10^{-5}$ & $-2.5$ & $-3.1$ & $0.10$ & $1.8\cdot10^{10}$ &
 $1.9\cdot10^{10}$ \\
$4.4\cdot10^{7}$ & $37.9$ & $1.8\cdot10^{4}$ & $2.0\cdot10^{11}$ &
 $8.4\cdot10^{-5}$ & $-2.4$ & $-3.0$ & $0.14$ & $1.9\cdot10^{10}$ &
 $2.0\cdot10^{10}$ \\
$6.0\cdot10^{7}$ & $39.5$ & $1.7\cdot10^{4}$ & $2.0\cdot10^{11}$ &
 $1.2\cdot10^{-4}$ & $-2.3$ & $-2.8$ & $0.20$ & $2.0\cdot10^{10}$ &
 $2.2\cdot10^{10}$ \\
$7.3\cdot10^{7}$ & $40.9$ & $1.6\cdot10^{4}$ & $2.0\cdot10^{11}$ &
 $1.5\cdot10^{-4}$ & $-2.2$ & $-2.7$ & $0.25$ & $2.0\cdot10^{10}$ &
 $2.3\cdot10^{10}$ \\
$9.8\cdot10^{7}$ & $43.8$ & $1.5\cdot10^{4}$ & $2.0\cdot10^{11}$ &
 $2.2\cdot10^{-4}$ & $-2.0$ & $-2.5$ & $0.36$ & $2.2\cdot10^{10}$ &
 $2.6\cdot10^{10}$ \\
$1.2\cdot10^{8}$ & $47.0$ & $1.4\cdot10^{4}$ & $2.0\cdot10^{11}$ &
 $3.0\cdot10^{-4}$ & $-1.9$ & $-2.4$ & $0.47$ & $2.3\cdot10^{10}$ &
 $2.9\cdot10^{10}$ \\
$1.7\cdot10^{8}$ & $54.6$ & $1.2\cdot10^{4}$ & $1.9\cdot10^{11}$ &
 $4.4\cdot10^{-4}$ & $-1.7$ & $-2.2$ & $0.72$ & $2.5\cdot10^{10}$ &
 $3.5\cdot10^{10}$ \\
$2.5\cdot10^{8}$ & $69.7$ & $8.7\cdot10^{3}$ & $1.9\cdot10^{11}$ &
 $7.2\cdot10^{-4}$ & $-1.5$ & $-2.0$ & $1.1$ & $2.7\cdot10^{10}$ &
 $4.5\cdot10^{10}$ \\
$3.0\cdot10^{8}$ & $83.3$ & $7.0\cdot10^{3}$ & $1.9\cdot10^{11}$ &
 $9.5\cdot10^{-4}$ & $-1.4$ & $-1.8$ & $1.5$ & $2.9\cdot10^{10}$ &
 $5.5\cdot10^{10}$ \\
$4.0\cdot10^{8}$ & $123.7$ & $4.3\cdot10^{3}$ & $1.8\cdot10^{11}$
&
 $1.6\cdot10^{-3}$ & $-1.2$ & $-1.6$ & $2.3$ & $3.2\cdot10^{10}$ &
 $8.1\cdot10^{10}$ \\
$5.6\cdot10^{8}$ & $243.5$ & $1.6\cdot10^{3}$ & $1.5\cdot10^{11}$
&
 $3.1\cdot10^{-3}$ & $-0.9$ & $-1.3$ & $4.0$ & $3.6\cdot10^{10}$ &
 $1.5\cdot10^{11}$ \\
$7.1\cdot10^{8}$ & $342.7$ & $6.2\cdot10^{2}$ & $1.1\cdot10^{11}$
&
 $6.9\cdot10^{-3}$ & $-0.5$ & $-1.0$ & $6.5$ & $3.8\cdot10^{10}$ &
 $2.5\cdot10^{11}$ \\
$9.2\cdot10^{8}$ & $161.7$ & $5.5\cdot10^{2}$ & $7.2\cdot10^{10}$
&
 $1.4\cdot10^{-2}$ & $-0.2$ & $-0.6$ & $8.6$ & $2.4\cdot10^{10}$ &
 $2.0\cdot10^{11}$ \\
$1.2\cdot10^{9}$ & $76.1$ & $4.9\cdot10^{2}$ & $4.7\cdot10^{10}$ &
 $2.1\cdot10^{-2}$ & $0.0$ & $-0.4$ & $8.2$ & $1.6\cdot10^{10}$ &
 $1.3\cdot10^{11}$ \\
$1.6\cdot10^{9}$ & $33.7$ & $5.4\cdot10^{2}$ & $3.3\cdot10^{10}$ &
 $2.7\cdot10^{-2}$ & $0.0$ & $0.0$ & $7.4$ & $1.2\cdot10^{10}$ &
 $8.9\cdot10^{10}$ \\
$2.1\cdot10^{9}$ & $21.1$ & $5.0\cdot10^{2}$ & $2.5\cdot10^{10}$ &
 $3.3\cdot10^{-2}$ & $0.1$ & $0.3$ & $6.8$ & $9.8\cdot10^{9}$ &
 $6.7\cdot10^{10}$ \\
$2.9\cdot10^{9}$ & $12.8$ & $4.1\cdot10^{2}$ & $1.8\cdot10^{10}$ &
 $3.8\cdot10^{-2}$ & $0.1$ & $0.4$ & $5.6$ & $9.5\cdot10^{9}$ &
 $5.3\cdot10^{10}$ \\
$3.8\cdot10^{9}$ & $7.2$ & $3.8\cdot10^{2}$ & $1.3\cdot10^{10}$ &
 $4.2\cdot10^{-2}$ & $0.2$ & $0.4$ & $4.4$ & $9.4\cdot10^{9}$ &
 $4.2\cdot10^{10}$ \\
    \hline
  \end{tabular}
\end{table*}

In the standard model, we adopted the value of [Fe/H] as a function of the
disk semi-thickness for the dependence of the iron abundance on the height
above the Galactic plane. This dependence is shown in Fig.~\ref{fig:feh-z},
along with data from \cite{Du04:AJ-128-2265}. We can distinguish three regions
with different metallicity gradients: $d{\textrm{[Fe/H]}}/dz = -3$/kpc for $z <
0.7$~kpc, $d{\textrm{[Fe/H]}}/dz =-0.2$/kpc for $0.7 < z < 5$~kpc, and
$d{\textrm{[Fe/H]}}/dz = -0.1$/kpc for $5 < z < 16$~kpc. This
corresponds to different iron abundances in stars of the thin disk, thick disk,
and halo, and also to different spatial scales within these components.

Computations of galactic evolution allowing for the loss of heavy elements
(Fig.~\ref{fig:feh-z}) and accretion have shown that the ``openness'' of the
galaxy weakly influences the model dependence of [Fe/H] on $z$. Thus, our
one-zone model can explain well the variations of the distribution of heavy
elements with $z$.

\section{Evolution of a disk galaxy: the influence of dark matter}

\begin{figure*}
  \psfrag{lg SFR}{$\lg$ SFR ($M_\odot/$yr)}
  \psfrag{lg H_g}{$\lg H_g$ (pc)}
  \psfrag{lg (M_g/M_G)}{$\lg(M_g/M_{\rm G})$}
  \psfrag{lg Z}{$\lg Z$}
  \psfrag{lg (tau_dust)}{$\lg \tau_{\rm dust}$}
  \psfrag{lg L_opt}{$\lg(L_{\rm opt}/L_\odot)$}
  \includegraphics[height=20cm]{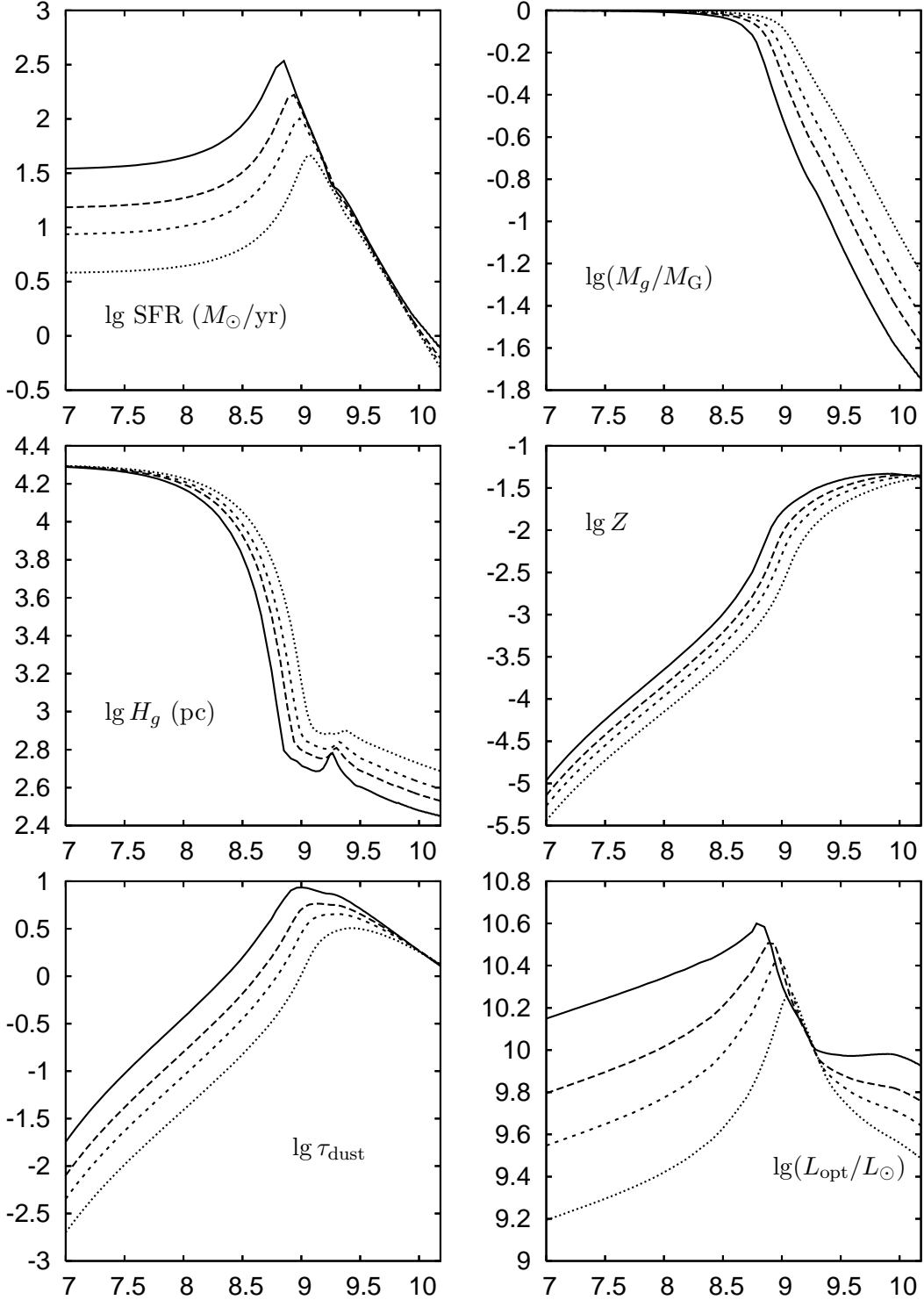}
  \caption{Evolution of the Galaxy with allowance for dark matter. SFR is the
    star-formation rate, $M_{\textrm{g}}/M_{\textrm{G}}$ the ratio of the
    masses of gas and visible matter, $H_{\textrm{g}}$ the semi-thickness of
    the gaseous disk, $Z$ the heavy-element abundance, $\tau_{\textrm{dust}}$
    the optical depth of the dust, and $L_{\textrm{opt}}$ the optical
    luminosity with allowance for absorption by dust. Computational results are
    shown for ratios of the masses of dark and visible matter
    $M_{\textrm{dm}}/M_{\textrm{G}}$ equal to 0 (solid curve), 0.5 (long dashed
    curve), 1.0 (short-dashed curve), and 2.0 (dotted curve).
    \hfill}
  \label{fig:Dark_Matter}
\end{figure*}

The main results obtained for the standard model are listed in the table.
Comparison with observations indicates that this model provides an
adequate description of the chemical and dynamical evolution of the Galaxy,
as well as of its star-formation history. For example, the data of
\cite{Liang04:AAP-423-867} provide evidence that young galaxies with
$\dot{M}_{*} \approx 10\text{--}100M_{\odot}$/yr have optical depths of $\sim
5\text{--} 10$ in the direction perpendicular to the plane of the disk, as is
confirmed by our model. The large optical depths of distant, and thus young,
galaxies could influence the extragalactic distance scale (which is based on
the assumption that SN~Ia are standard candles), and therefore conclusions
concerning the regime of the cosmological expansion. We should also bear this
in mind in connection with analyses of the brightness of supernovae at $z =
2\text{--}5$. Variation of the time to accumulate the mass of the Galaxy in the
initial stage of its formation did not influence substantially the Galaxy's
parameters after the Hubble time (Fig.~\ref{fig:ofe-feh}).

Our model has another important parameter --- the mass of dark matter,
i.e., of matter (baryonic or otherwise) that cannot be directly observed, but
provides a (sometimes dominant) contribution to the gravitational field.
The fractional mass of dark matter is small on small scales. For instance,
estimates of the fractional mass of dark matter in the Sun yield less than
2--5\% \cite{Kardashov05:AZ-81}, and this is probably true for other stars as
well. On scales of 100~kpc or more, the fractional mass of dark matter may be
much larger \cite{Shustov05:ISSC-207}.

We studied the role of dark matter in the evolution of the Galaxy by changing
the expression for the gravitational potential in the standard model to
\begin{gather*}
  E = \frac{G M_{\textrm{g}} H_{\textrm{g}}^2}{R^2}
  \left( \frac{M_{\textrm{G}}}{H_{*}} + \frac{M_{\textrm{dm}}}{R} \right),
\end{gather*}
where $G$ is the gravitational constant, $M_{\textrm{g}}$ the mass of gas,
$H_{\textrm{g}}$ the semi-thickness of the gaseous disk, $M_{\textrm{G}}$ the
mass of visible matter, $H_{*}$ the semi-thickness of the stellar disk,
$M_{\textrm{dm}}$ the mass of dark matter, and $R$ the radius of the disk.
Figure~\ref{fig:Dark_Matter} shows the evolution of the main parameters of the
Galaxy as functions of the ratio of the masses of dark and visible matter,
which was varied from zero (corresponding to the standard model) to
two. Reduction of the mass of matter involved in star formation results in a
delay of the star-formation burst, and so reduces the thickness of the gaseous
disk. This is in qualitative agreement with estimates following from the
observational relation between the relative thickness of the disk and the
mass--luminosity relation for galaxies \cite{Zasov02:ASTL-28-527}. Increasing
the mass-to-luminosity ratio in the model decreases the galaxy's relative
thickness. The observed optical luminosity of the Galaxy places constraints on
the fractional mass of dark matter, $<50$\%. Note that our initial mass
function, $dN \propto M^{-2.35}dM$ for $M > 0.1~M_{\odot}$, overestimates the
number of low-mass stars compared to the observed value. According to
\cite{Kroupa02:MNRAS-336-1188}, the observed mass function changes slope at $M
\approx 1M_{\odot}$ ($dN \propto M^{-1.5}dM$ for $M < 1 M_{\odot}$), which
makes the fraction of stars with $< 1 M_{\odot}$ less pronounced. Thus, an
excess of matter appears in our model compared to the observations. A simple
analytical estimate of its fractional mass is $\sim 37$\%.

\section{Tully--Fisher relation}

\begin{figure*}[t!]
  \psfrag{lg(R)}{$\lg R$ (pc)}
  \psfrag{lg(M)}{$\lg(M/M_\odot)$}
  \psfrag{tau_SF=tau_Hubble}{$\tau_{\rm SF} = \tau_{\rm H}$}
  \psfrag{tau_SF=tau_SN}{$\tau_{\rm SF} = \tau_{\rm SN}$}
  \psfrag{LSBG}{LSBG}
  \psfrag{T-F}{T-F}
  \psfrag{EG}{EG}
  \includegraphics[width=15cm]{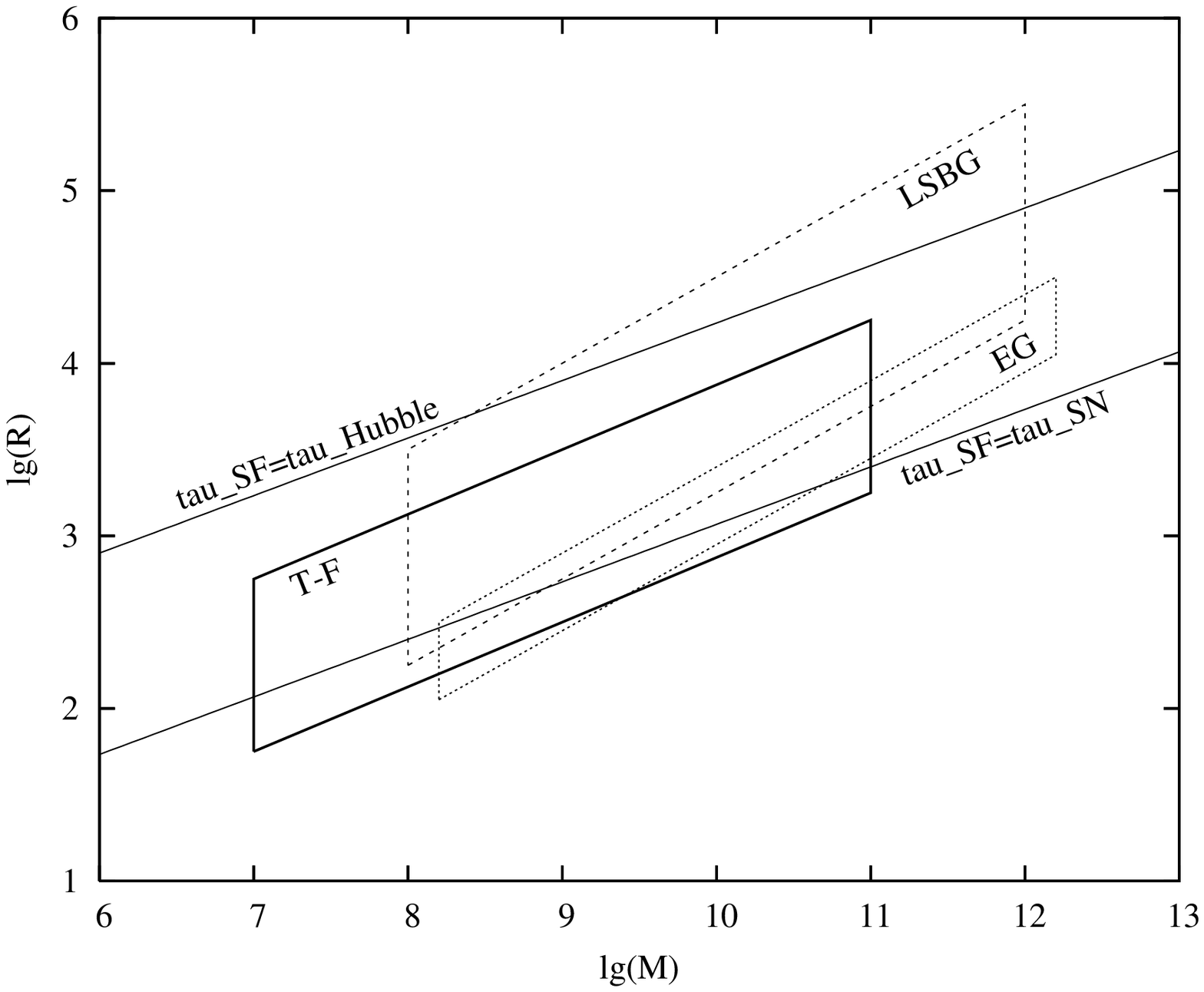}
  \caption{Mass--radius diagram for various galaxies; galaxies obey the
    Tully--Fisher relation in the region T--F. The lines $\tau_{\textrm{SF}} =
    \tau_{\textrm{SN}}$ and $\tau_{\textrm{SF}} = \tau_{\textrm{Hubble}}$
    define the positions of galaxies with star-formation timescales
    $\tau_{\textrm{SF}}$ and $\tau_{\textrm{H}}$, respectively. The region LSBG
    is occupied by low-surface-brightness galaxies
    \cite{Kniazev04:AJ-127-704,Okoshi:astro-ph:0412561}. Early-type galaxies
    are located in the region EG
    \cite{Rijcke:astro-ph:0412553,Walcher:astro-ph:0409216} (see Sect. 5).
    \hfill}
  \label{fig:Tully-Fisher_Stripe}
\end{figure*}

About a quarter of a century ago, Tully and Fisher \cite{Tully77:AAP-54-661}
discovered a dependence between the luminosity of a disk galaxy and its
rotational velocity.  In modern form, this relation can be written
\begin{gather}
  \Delta M_{\textrm{B}} = (8 \pm 1)\Delta \log v_{\textrm{rot}},
  \label{eq:M_B-v_rot}
\end{gather}
where $M_{\textrm{B}}$ is the absolute $B$ magnitude of the galaxy and
$v_{\textrm{rot}}$ is the rotational velocity of the galactic disk, estimated
from the flat part of the rotation curve. We can use the mass--luminosity
relation for disk galaxies, $M/M_\odot \sim 10 L/L_\odot$
\cite{Karachentsev:astro-ph:0412369}, to estimate the relation between the mass
of a disk galaxy $M$ and its radius $R$. Assuming $v_{\textrm{rot}} \propto
M^{1/2} R^{-1/2}$, we obtain
\begin{equation}
  R \propto M^{0.5 \pm 0.07}  \;.
  \label{eq:Tully-Fisher}
\end{equation}
Relations similar to (\ref{eq:M_B-v_rot}) for varioius groups of disk galaxies
have been studied by numerous authors (see, e. g.,
\cite{Kannappan04:AJ-127-2694,Pilyugin04:AAP-425-849,Andersen03:APJL-599-L79}).
If selection effects are not considered, it appears that the locations and
slopes of the $\log R\text{--}log M$ relation derived from (\ref{eq:M_B-v_rot})
are fairly uncertain (they vary by about a factor of three for galaxies with
masses from $10^7$ to $10^{11}M_{\odot}$). The Tully--Fisher relation can be
written $M = \pi \Sigma R^2$, where $\Sigma$ is the surface density of
gravitating matter. The observed range of $\Sigma$ is rather wide. According
to \cite{Kniazev04:AJ-127-704}, disk galaxies display values $\Sigma \approx
10\text{--}600~M_{\odot}/\textrm{pc}^2$; another estimate yields $\Sigma =
20\text{--}1500 M_{\odot}/\textrm{pc}^2$
\cite{Pfenniger04:ESA-DMU-63}. Figure~\ref{fig:Tully-Fisher_Stripe} plots a
mass--radius diagram for galaxies of various types that obey the Tully--Fisher
relation (located inside the ``T--F'' region in this figure), elliptical and
early-type galaxies (EG;
\cite{Rijcke:astro-ph:0412553,Walcher:astro-ph:0409216}), and
low-surface-density galaxies (LSBG;
\cite{Kniazev04:AJ-127-704,Okoshi:astro-ph:0412561}).

We used (\ref{eq:Tully-Fisher}) to compute a series of models with masses of
$10^6 - 4.5\times10^{11}M_{\odot}$. The results show that, in this range, the
mass of a galaxy exerts almost no influence on the star-formation history or
the evolution of the chemical composition. In particular, the final
heavy-element abundance does not depend on the mass, in contradiction with the
observed increase of the heavy-element abundance with increasing galactic mass
and luminosity \cite{Salzer:astro-ph:0502202}. This may be explained by the
loss of heavy elements by low-mass galaxies.

Our formalism \cite{Firmani92:AAP-264-37,Firmani94:AAP-288-713} enables us to
derive a simple relation defining the time scale for star formation in a
spherical galaxy \cite{Tutukov03:RMXAC-17-60}:
\begin{gather*}
  \tau_{\textrm{SF}} = 3\times10^8 \frac{R_4^3}{M_{11}},
\end{gather*}
where $R_4$ is the galactic radius in units of $10^4$~pc and $M_{11}$ the mass
of its gaseous component in units of $10^{11}M_{\odot}$. The observed masses
and radii of young galaxies are consistent with $\tau_{\textrm{SF}} =
10^7\text{--}10^9$~yr \cite{Papovich:astro-ph:0501088}. The spherical model
describes the early stages of the evolution of disk galaxies. We can place
limits on the time for the existence of star-forming galaxies. An upper bound
for the region they can occupy in Fig.~4 is given by the Hubble time:
$\tau_{\textrm{SF}}  = \tau_{\textrm{H}} = 1.4\times10^{10}$~yr, as is shown in
Fig.~\ref{fig:Tully-Fisher_Stripe}. In galaxies located above this line, the
SFR is so low that only a small fraction of the gas has been turned into stars
over the Hubble time.  Note that the position of this bound virtually coincides
with the position occupied by low-surface-brightness galaxies
\cite{Kniazev04:AJ-127-704,Okoshi:astro-ph:0412561}. Nearby galaxies with
masses of $10^6 \text{--} 10^{12}M_{\odot}$ from the catalog
\cite{Karachentsev04:AJ-127-2068} abut on the upper boundary of the region of
star-forming galaxies (they are not shown in
Fig.~\ref{fig:Tully-Fisher_Stripe}); the star-formation time scale in such
galaxies is $\tau_{\textrm{SF}} \approx 80^9 \text{--}10^{10}$~yr.

On the other hand, the SFR in high-density galaxies is so high that almost
all the gas has been transformed into stars even before the first explosions
of SN~II, whose precursors have lifetimes $\tau_{\textrm{SN}} \approx
5\times10^6$~yr.  This corresponds to the relation  $\tau_{\textrm{SN}} <
\tau_{\textrm{SF}}$, or $M_{11} > 60 R_4^3$ (the line $\tau_{\textrm{SN}} =
\tau_{\textrm{SF}}$ is also plotted in Fig.~\ref{fig:Tully-Fisher_Stripe}).  As
soon as the SN~II begin to explode in these galaxies, they clean out any
remaining gas. This is the scenario for the evolution for elliptical
galaxies. The mass lost by old stars in these galaxies is unable to revive star
formation, since SN~Ia explosions drive the galactic wind, hindering star
formation throughout the galaxy, with the possible exception of a circumnuclear
region with a high gas density (see, for example,
\cite{Chilingarien:astro-ph:0412293}). In fact, only a few percent of S0 and E
galaxies show signs of star formation in their nuclei
\cite{Fukugita04:APJL-601-L127}.  Some S0 and E galaxies may form via
collisions of disk galaxies
\cite{Keel03:AJ-126-1257,Nipoti:astro-ph:0311424}. However, the number of such
galaxies is not large, since observations \cite{Aguerri02:MNRAS-333-633} show
only a small increase in the fraction of E galaxies with the age of Universe;
hence, most formed at large redshifts. Albeit with a fairly large scatter, the
positions of early-type galaxies in Fig.~\ref{fig:Tully-Fisher_Stripe} are
close to the boundary defined by the relation $\tau_{\textrm{SF}} =
\tau_{\textrm{SN}}$. However, elliptical galaxies formed in collisions of disk
galaxies may also have larger sizes.

Figure~\ref{fig:Tully-Fisher_Stripe} shows that the galaxies with active
continuing star formation are located in a strip defined by the conditions
$\tau_{\textrm{SN}} = \tau_{\textrm{SF}}$ and $\tau_{\textrm{SN}} =
\tau_{\textrm{H}}$; the slope of this strip is close to the slope of the
Tully--Fisher relation. This suggests that the power of the $R \propto
M^\alpha$ law is determined by the similarity of the slopes of the boundaries
of the region occupied by disk galaxies in the mass--radius diagram.  At the
lower boundary of this zone, galaxies that are initially elliptical are formed,
while the upper boundary is apparently defined by two circumstances. The first
is the condition that there be a sufficient SFR on a time scale that is shorter
than the Hubble time. The second is the near coincidence of the upper boundary
with the position of galaxies that have surface brightnesses at the detection
limit. Galaxies with lower surface densities, $\Sigma < 10
M_{\odot}/\textrm{pc}^2$, are probably not detectable.

\section{Conclusion}

Our study of the evolution of a disk galaxy with star formation governed by
ionization has enabled us to apply this model to several new fields. In
particular, we have shown that high oxygen-to-iron abundance ratios are
characteristic of the very first stars formed: $\textrm{[O/Fe]} \approx 2$
when $\textrm{[Fe/H]} \lesssim -5$. This is a result of the reduction in
the iron production and increase in the oxygen production with increasing
initial mass of type~II supernovae.  This suggests that some stars with
low iron abundances and relatively high oxygen abundances (G~77-61,
HE~0107-5240, Fig.~\ref{fig:ofe-feh}) may be either the earliest
second-generation stars or first-generation stars that are ``contaminated'' by
products of the first supernovae. Our model distribution of metals with height
$z$ above the Galactic disk agrees with the observed distribution to $z \sim
16$~kpc. In our model, the well-known Tully--Fisher relation may stem from a
combination of observational selection effects and conditions that are
necessary for the formation of disk galaxies.

\section{Acknowledgments}

This study was supported by the Presidential Program ``Leading Scientific
Schools of Russia'' (NSh-162.2003.2), the Russian Foundation for Basic
Research (project codes 02-02-17524, 03-02-162254), and the Program of the
Physical Sciences Branch of the Russian Academy of Sciences ``Extended
Objects in the Universe.''

\newpage
\bibliography{rep}

\end{document}